\documentclass[aps,prl,final,twocolumn,superscriptaddress,floatfix]{revtex4-1}

\usepackage[latin9]{inputenc}
\usepackage{graphicx,amssymb,soul,color,amsmath,bm}

\begin{document}

\title{Diverging Nematic Susceptibility, Physical Meaning of $T^*$ scale, \\
and Pseudogap in the Spin Fermion Model for Pnictides}

\author{Shuhua Liang}
\affiliation{Materials Science and Technology Division, Oak Ridge National Laboratory, Oak Ridge, Tennessee 37831, USA}
\affiliation{Department of Physics and Astronomy, The University of Tennessee, Knoxville, Tennessee 37996, USA}

\author{Anamitra Mukherjee}
\affiliation{Department of Physics and Astronomy, The University of Tennessee, Knoxville, Tennessee 37996, USA}

\author{Niravkumar D. Patel}
\affiliation{Department of Physics and Astronomy, The University of Tennessee, Knoxville, Tennessee 37996, USA}

\author{Elbio Dagotto}
\affiliation{Materials Science and Technology Division, Oak Ridge National Laboratory, Oak Ridge, Tennessee 37831, USA}
\affiliation{Department of Physics and Astronomy, The University of Tennessee, Knoxville, Tennessee 37996, USA}

\author{Adriana Moreo}
\affiliation{Materials Science and Technology Division, Oak Ridge National Laboratory, Oak Ridge, Tennessee 37831, USA}
\affiliation{Department of Physics and Astronomy, The University of Tennessee, Knoxville, Tennessee 37996, USA}

\date{\today}

\begin{abstract}
Using Monte Carlo simulations with a tunable uniaxial strain, 
for the first time the nematic susceptibility of a model for the pnictides 
is calculated. The results are in good agreement with the
experiments by J-H. Chu {\it et al.}, Science {\bf 337}, 710 (2012). 
Via a Ginzburg-Landau analysis, our study
suggests a nematicity in pnictides primarily originating
on magnetism, but with the lattice/orbital boosting up critical temperatures and
separating the structural $T_S$ and N\'eel $T_N$ transitions. 
At $T>T_S$, Curie-Weiss behavior is observed with the characteristic temperature 
$T^*$ unveiled by Chu {\it et al.} being the $T_N$ of the purely 
electronic system. In  this temperature regime,
short-range magnetic order with wavevectors $(\pi,0)-(0,\pi)$ 
induce local nematic fluctuations and a density-of-states pseudogap, 
compatible with several experiments.
\end{abstract}


\maketitle

{\it Introduction.} The complexity of high critical 
temperature iron-based superconductors~\cite{johnston,dai}, with
coupled spin, charge, orbital, and lattice degrees of freedom (DOF), 
creates exotic regimes such as the widely discussed
nematic state with broken rotational invariance~\cite{fisher,fradkin}.
This state may originate in the spin DOF~\cite{spin3,spin1,spin2,fernandes1,fernandes2}
or in the orbital DOF~\cite{bnl2,orb1,orb2,kontani}, but
subtleties in experiments (with strain required to detwin crystals)
and in theory (employing complicated multiorbital models)
have prevented the identification of the primary driver 
of the nematic regime.

Recent efforts to study nematicity have considered models 
with electrons coupled to the lattice~\cite{PRL-2013}. The electronic sector 
is itself separated into itinerant and localized electrons defining a 
spin-fermion (SF) model~\cite{BNL,kruger,PRL-2012,CMR}, compatible 
with the growing evidence that iron-superconductors display a
mixture of itinerant and localized features~\cite{dai,loca1,loca2}.
These studies unmasked 
a considerable electron-lattice feedback, leading to several results
in agreement with experiments, such as anisotropic resistivities and a
nematic and structural (tetragonal-orthorhombic) 
transition at $T_S$, slightly separated from 
the N\'eel temperature $T_N$ ($<T_S$)~\cite{wide}.

More recently,  a remarkable experimental
development has been the report of a diverging nematic
susceptibility $\chi^{exp}$ vs. temperature $T$, with a mysterious characteristic
temperature scale $T^*$, for
single crystals of Ba(Fe$_{1-x}$Co$_x$)$_2$As$_2$~\cite{fisher-science}
measured by varying an in-situ uniaxial strain.
Although contrasting $\chi^{exp}$ against theory and explaining
the physical meaning of $T^*$ 
are  crucial aspects to identify the mechanism that drives nematicity, 
to our knowledge $\chi^{exp}$ and $T^*$ have not 
been addressed theoretically before
since temperatures above $T_S$ are difficult to
study with reliable methods.

In this publication, for the first time this nematic susceptibility 
is theoretically calculated via the SF model coupled to the lattice 
in precisely the same setup as in~\cite{fisher-science}. Note that this
susceptibility, that tests a local geometric property of an enlarged
parameter space, is {\it different} from the simpler magnetic susceptibility
calculated in~\cite{PRL-2013} obtained from thermal statistics.
The present computational effort required an order 
of magnitude more work than in~\cite{PRL-2013} 
because the strain is an extra parameter to vary, 
rather than being dynamically
adjusted in the Monte Carlo (MC) process as before. To implement this
demanding task, modifications in the MC algorithm were implemented,
as explained below.
Compared to Hubbard multiorbital approaches, 
a unique characteristic of the SF model is 
that simulations can be carried out 
in the nematic regime above the ordering temperatures.
Remarkably, our susceptibility is very similar to the
diverging experimental $\chi^{exp}$ result. Moreover,
we observed that the $T^*$ scale in the Curie-Weiss behavior 
is the preexisting magnetic critical temperature of the
purely electronic sector, which is independent of the lattice. 
We also observed a density-of-states pseudogap and nematic fluctuations above $T_S$, 
caused by short-range $(\pi,0)$-$(0,\pi)$ antiferromagnetic order.

{\it Models.} The model employed here combines
the purely electronic spin-fermion model~\cite{BNL,kruger,PRL-2012,CMR} 
together with lattice orthorrombic distortions:
\begin{equation}
H_{\rm SF} = H_{\rm Hopp} + H_{\rm Hund} + H_{\rm Heis} + H_{\rm SL} + H_{\rm OL} + H_{\rm Stiff}.
\label{ham}
\end{equation}
This (lengthy) full Hamiltonian is in the Supplementary 
Material~\cite{SM}.  
$H_{\rm Hopp}$ is the Fe-Fe hopping of the $d_{xz}$, $d_{yz}$, 
and  $d_{xy}$ electrons (three orbitals model; bandwidth $W$$\sim$$3$~eV), with 
amplitudes that reproduce photoemission results. The average number of electrons 
per itinerant orbital is $n$=4/3~\cite{three} (undoped regime) since many
nematic-state experiments are carried out in this limit, and technically
the study simplifies in the absence of doping and quenched disorder.
The Hund interaction is canonical:
$H_{\rm Hund}$=$-{J_{\rm H}}\sum_{{\bf i},\alpha} {{{\bf S}_{\bf i}}\cdot{{\bf s}_{{\bf i},\alpha}}}$,
with ${{\bf S}_{\bf i}}$ (${\bf s}_{{\bf i},\alpha}$) 
the localized (itinerant with orbital index $\alpha$) spin. 
$H_{\rm Heis}$ is the Heisenberg interaction 
among the localized spins involving
nearest-neighbors (NN) and next-NN (NNN) interactions with couplings $J_{\rm NN}$
and $J_{\rm NNN}$, respectively, 
and ratio $J_{\rm NNN}$/$J_{\rm NN}$=2/3~\cite{PRL-2012} 
to favor collinear order.
Within the spin-driven scenario for nematicity, 
the state between $T_N$ and $T_S$ is characterized by short-range 
spin correlations $\Psi_{\bf i}$=$\sum_{\pm} ({{\bf{S}_{\bf i}}\cdot{\bf{S}_{{\bf i}\pm{\bf y}}}} 
- {{\bf{S}_{\bf i}}\cdot{\bf{S}_{{\bf i}\pm{\bf x}}}})/2$ that satisfy 
$\langle \Psi \rangle$$>$$0$~\cite{fernandes2,clari}, 
where ${\bf{S}_{\bf i}}$ is the spin
of the iron atom at site ${\bf i}$ and ${\bf x,y}$ are unit vectors along the axes.
The $\mathcal{O}_{rth}$-distortion $\epsilon_{\bf i}$
associated to the elastic constant $c_{66}$ 
will be considered here~\cite{SM}. 
The coupling of the spin-nematic order and the lattice is 
$H_{\rm SL}$=$-g\sum_{\bf i}\Psi_{\bf i}\epsilon_{\bf i}$~\cite{fernandes1,fernandes2}, 
where $g$ is the lattice-spin coupling~\cite{only}.
To also incorporate orbital fluctuations, the term 
$H_{\rm OL}$=$-\lambda\sum_{\bf i}\Phi_{\bf i}\epsilon_{\bf i}$ is added, where
$\lambda$ is the orbital-lattice coupling, 
$\Phi_{\bf i}$=$n_{{\bf i},xz}$-$n_{{\bf i},yz}$ is the orbital 
order parameter, and $n_{{\bf i},\alpha}$ the electronic density at site
${\bf i}$ and orbital $\alpha$~\cite{kontani}.
Finally, $H_{\rm Stiff}$ is the spin stiffness given by 
a Lennard-Jones potential that speeds up convergence~\cite{SM}.


{\it Many-body techniques.} 
The Monte Carlo method used in this study is well known~\cite{PRL-2012,CMR}, 
and details will not be repeated. However, here
an extra computational component 
had to be introduced because, compared with~\cite{PRL-2013},
for each temperature $T$ now the strain was varied as an extra parameter. Since for each $T$
typically 15 values of strain were used, this effort is $\sim$15 times more costly than
in~\cite{PRL-2013}.
While 
the standard Monte Carlo is time consuming because of the fermionic-sector 
exact diagonalization (ED) at every step, in the related 
double-exchange models for manganites an improvement has
been used before: the
``Traveling Cluster Approximation'' (TCA)~\cite{kumar} where the MC updates 
are decided employing a cluster centered 
at site ${\bf i}$ with a size substantially smaller than the full lattice size~\cite{steps}. 
In addition, twisted boundary 
conditions (TBC) were also used~\cite{salafranca}. This is the first time
that TCA and TBC are employed together.
To simplify further the analysis, most couplings are fixed to values used
successfully before~\cite{PRL-2012}: ${J_{\rm H}}$=$0.1$~eV, 
${J_{\rm NN}}$=$0.012$~eV, and ${J_{\rm NNN}}$=$0.008$~eV. 
The dimensionless versions of the electron-phonon couplings $\tilde g$ and
$\tilde \lambda$ are fixed to 0.16 and 0.12, respectively, as in~\cite{PRL-2013}, 
although results for other values can be found in~\cite{SM}.

\begin{figure}[thbp]
\begin{center}
\includegraphics[trim = 7mm 0mm -7mm 0mm,width=0.43\textwidth,angle=0]{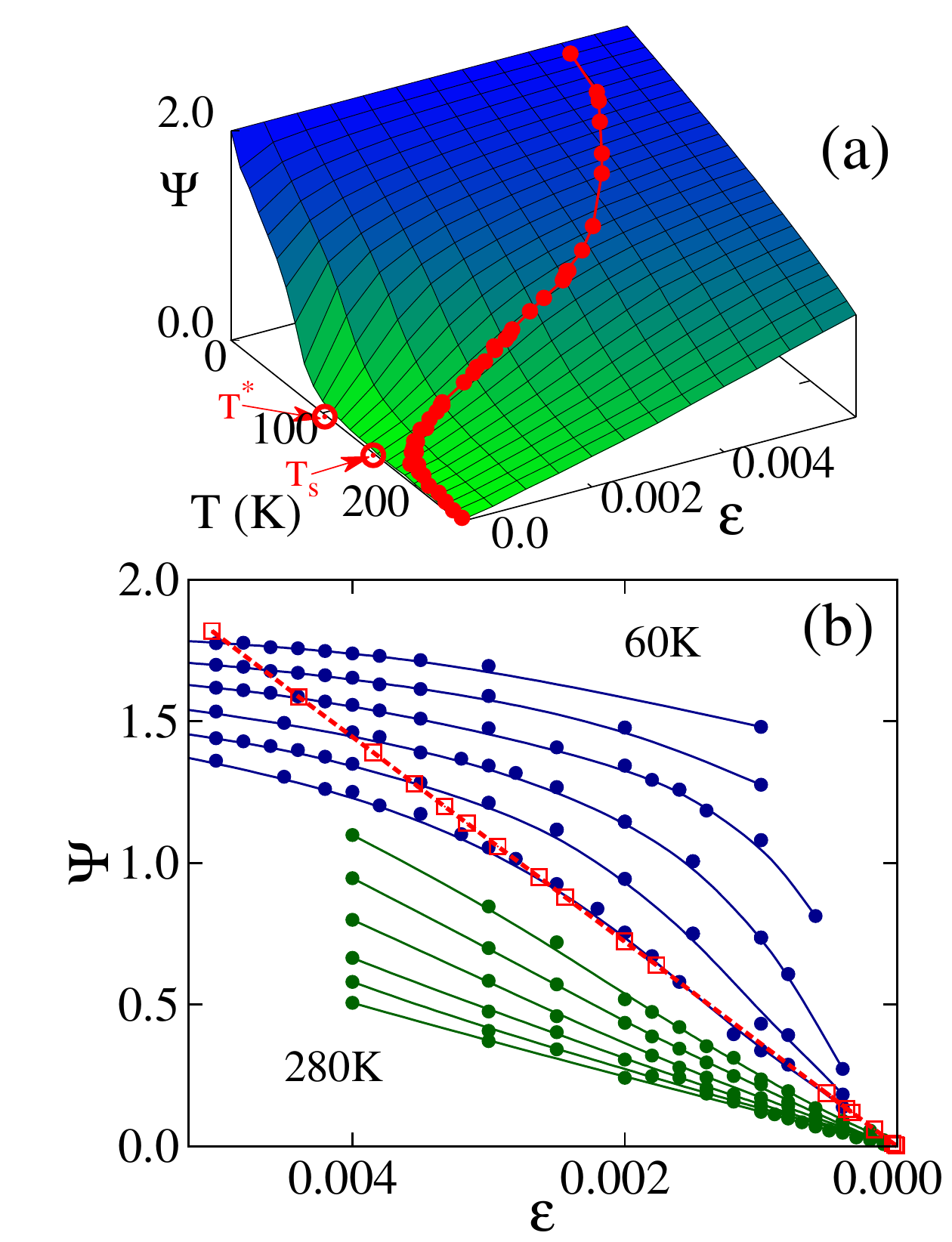}\label{newfiga}
\vskip -0.2cm
\caption{(color online)
Monte Carlo spin-nematic order parameter, 
at $\tilde g$=$0.16$ and $\tilde \lambda$=$0.12$.
(a) $\Psi$ vs. $T$ and $\epsilon$, 
measured at a fixed lattice distortion 
$\epsilon$ for each $T$ (restricted MC). Shown are the $T^*$ 
(see text) and $T_S$ ($\sim T_N$)  temperatures.
Data are for an 8$\times$8 cluster with TCA+TBC 
(PBC 8$\times$8 clusters with ED give similar results). Red points 
are the equilibrium values using unrestricted MC with ED and PBC 8$\times$8 clusters.
(b) $\Psi$ vs. $\epsilon$ at fixed temperatures, illustrating their nearly linear
relation in unrestricted MC (red), and also the
linear slopes of the restricted MC curves (green/blue) close to $T_S$.
Results are obtained with ED/PBC 8$\times$8 clusters.
Note that the number of green/blue points vastly outnumbers the number of red points,
highlighting how much more demanding this effort has been than in~\cite{PRL-2013}.}
\vskip-0.6cm
\label{newfig}
\end{center}
\end{figure}

The spin nematic susceptibility calculated here is defined as
$\chi_s=\frac{\partial \Psi}{\partial \epsilon}|_{\epsilon_0}$
where $\epsilon_0$ is the value of the lattice distortion 
obtained from the ``unrestricted'' numerical simulation where
the lattice is equilibrated together with the spins, as in~\cite{PRL-2013}.
To calculate $\chi_s$ of our model, 
a procedure similar to the experimental setup was employed: the order
parameter $\Psi$ was measured at various temperatures and at {\it fixed}
values of the lattice distortion $\epsilon$=$(a_x-a_y)/(a_x+a_y)$ (``restricted'' MC). 
By this procedure,
$\Psi({\tilde g},{\tilde \lambda},T,\epsilon)$ are obtained at fixed couplings, 
defining surfaces as in Fig.~\ref{newfig}(a). Allowing the lattice
to relax the equilibrium curve
[red, Fig.~\ref{newfig}(a)] is obtained.

\vskip -0.1cm

Figure~\ref{newfig}(b) contains the (restricted) MC measured spin-nematic order
parameter versus the (fixed) lattice distortion $\epsilon$, at various temperatures. 
In a wide range of temperatures, a robust linear behavior is observed 
and $\chi_s$ can be easily extracted numerically. Figure~\ref{newfig}(b)
is similar to the experimental
results in Fig.~2A of Ref.~\cite{fisher-science}.
The equilibrium result with both spins and lattice optimized 
(unrestricted MC) is also shown (red squares).

\begin{figure}[thbp]
\includegraphics[trim = 7mm 0mm -7mm 0mm,width=0.55\textwidth,angle=0]{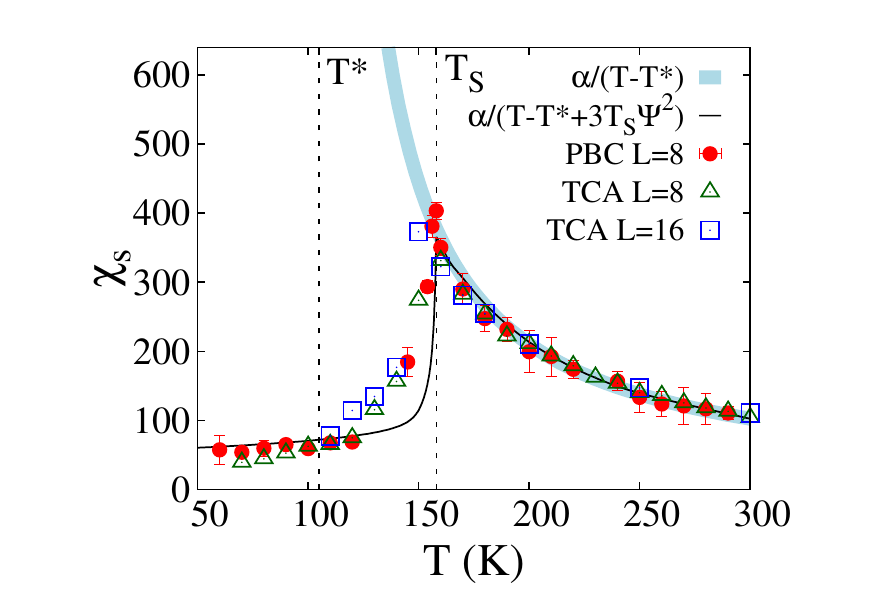}\label{fig:1a}
\vskip -0.3cm
\caption{(color online) Nematic susceptibility $\chi_s$ of the
SF model
vs. temperature $T$ (circles, triangles, and squares) 
obtained from Fig.~\ref{newfig}(b), at the
realistic couplings $\tilde g$=$0.16$ and $\tilde \lambda$=$0.12$ 
($\alpha$=$\tilde g$/$a_0$). 
Two MC techniques were employed: 
``PBC L=8'' is the standard MC method 
with ED in the fermions 
at every step, using 8$\times$8 clusters with PBC.
``TCA L=8'' and ``TCA L=16'' correspond to the TCA+TBC method 
on L$\times$L clusters. Size effects are small. 
Also shown are two GL fits: the light blue (thick)
line displays the Curie-Weiss 
equation $\chi_s\approx {\tilde g\over{a_0(T-T^*)}}$, 
indicating a divergence at a lower temperature $T^*$, 
characteristic of the electronic sector alone.
At $T \leq T_S$, the lattice follows the electronic sector. The
black (thin) line is Eq.(\ref{chisfit}) with the
$3T_S\Psi^2$ correction (see text)~\cite{T8T16}. 
}
\vskip-0.4cm
\label{fig:1}
\end{figure}

Our main result is presented in Fig.~\ref{fig:1}, where 
the numerically calculated $\chi_s$ vs. $T$ is displayed, at the 
realistic couplings used in
previous investigations~\cite{PRL-2013}. 
In remarkable agreement with experiments,
$\chi_s$ grows when cooling down and it develops 
a sharp peak at $T_S$ (compare with Fig.~2B of Ref.~\cite{fisher-science}).
These results were obtained via two different procedures 
(standard ED and the TCA+TBC), 
and for two lattice sizes, indicating that systematic 
errors (such as size effects) are small.

{\it Analysis of $\chi_s$ results.} 
Supplementing the computational results, here Ginzburg-Landau (GL) 
calculations were also performed, 
similarly as in~\cite{fisher-science} for experiments. 
Note that the previous GL analysis considered only a generic
nematic order parameter while our study separates the spin and orbital
contributions. The rather complex 
numerical results can be rationalized quantitatively by this procedure.
The results for $\chi_s$ (Fig.~\ref{fig:1}) are well fitted quantitatively 
for $T>T_S$, and qualitatively for $T<T_S$, by the expression:
\begin{equation}
\chi_s=\frac{\tilde g}{a_0[(T-T^*)+3T_S \Psi^2]},
\label{chisfit}
\end{equation}
\noindent where $T_S$=$158$~K, $T^*$=$105$~K, and $a_0$$\sim$$0.093$.  
$T^*$ and $a_0$ are here mere fitting parameters, but the GL
analysis~\cite{SM} shows that $a_0$ arises from the
GL quadratic term $a\Psi^2/2$ in a second order
transition where $a=a_0(T-T^*)$. 
$\Psi$ is the equilibrium value from unrestricted 
MC simulations [red, Fig.~\ref{newfig}(a)] and it is $T$ dependent. 
For $T\ge T_S$, $\Psi$ vanishes
and $\chi_s$  
exhibits Curie-Weiss behavior, in excellent agreement with the
experimental $\chi^{exp}$~\cite{fisher-science}.

Let us discuss the meaning of the fitting parameter $T^*$:

(1) From Fig.~\ref{newfig}(b), the unrestricted numerical results at $T=T_S$ indicate a
linear relation between $\Psi$ and $\epsilon$ 
(while individually both behave as order parameters, i.e. they change fast near $T_S$). 
$\epsilon=\epsilon(T)$ because 
the lattice is equilibrated together with the spins.
However, this nearly temperature independent ratio $\Psi/\epsilon$=$K$~($\sim$$360$)
depends on couplings: comparing results at several $\tilde g$s, it is
empirically concluded that $K={\hat c\over{\tilde g}}$ (constant $\hat c$). 

Note also that $\chi_s$  depends on the partial derivative 
${\partial\Psi /{\partial\epsilon}}|_{\epsilon_0}$, since
$\chi_s$ is obtained at a constant $T$ varying $\epsilon$ via strain to match the 
procedure followed in experiments~\cite{fisher-science}, in the vicinity
of the equilibrium point $\epsilon_0$ [$\chi_s$ arises
from the green/blue curves of Fig.~\ref{newfig}(b), not from the
red equilibrium curve]. While these slopes (restricted vs. unrestricted MC)
are in general different, both become very similar 
at $T \sim T_S$ where it can be shown analytically that these derivatives 
are indeed almost the same~\cite{SM}.
Thus, at $T_S$:
${d\Psi\over{d\epsilon}}=
{{\hat c}\over{\tilde g}}\approx{\partial\Psi\over{\partial\epsilon}}|_{\epsilon_0}=\chi_s$.
This relation can be independently 
deduced from the 
GL analysis, Eq.(\ref{linear}), with $\hat c$=$c_0$, and $c_0$ arising from 
$c_0\epsilon^2/2$ in the free energy, providing physical meaning to 
parameters in the MC fits.

(2) 
Since the numerical susceptibility $\chi_s$ can be fit well
by Eq.(\ref{chisfit}) including at $T_S$ where $\Psi = 0$, then
$T_S=T^*+{\tilde g^2\over{a_0{\hat c}}}$~\cite{fisher-science,kasahara,SM}.
Comparing with Eq.(\ref{tsa}), 
$\hat c$ is again identified with the uncoupled shear 
elastic modulus $c_0$.
In addition, from~\cite{PRL-2012} it is known 
that at $\tilde g$=$\tilde \lambda$=$0$ there is no nematic regime and 
$T_S$=$T_N$, the N\'eel temperature. Then,
$T_N=T^*+{\tilde g^2\over{a_0c_0}}$,
that at $\tilde g=0$ leads to the important conclusion that {\it the scale $T^*$ 
is simply equal to the N\'eel temperature of the purely electronic 
SF model}. In previous work~\cite{PRL-2012} it was reported 
that $T_N$ at $\tilde g$=$\tilde \lambda$=$0$ 
is $\sim$100-110~K, in remarkable agreement with the fitting value
of $T^*$ obtained {\it independently}. 
Thus, in the Curie-Weiss formula $T^*$ is solely 
determined by the magnetic properties of 
the purely electronic system. This suggests 
that the magnetic DOF in the SF model plays a leading role
 to explain the nematic state of 
Ba(Fe$_{1-x}$Co$_x$)$_2$As$_2$~\cite{fisher-science}.
However, the lattice/orbital DOS are also crucial to boost 
the critical temperature
from $T^*$$\sim$$105$~K to $T_S$$\sim$$158$~K~\cite{orbitalsusce}.
%


{\it $T_S$ vs. $\tilde \lambda$. } 
The study in Figs.~\ref{newfig}(a,b) was repeated 
for other $\tilde \lambda$s.
It was observed that $\hat c$ varies with $\tilde \lambda$, compatible with
the GL analysis where 
${\hat c}(\tilde\lambda)=c_0(1-{\tilde\lambda^2\over{e_0c_0}})$, Eq.(\ref{ceff}). 
At small $\tilde \lambda$, the total (unrestricted MC) 
and partial (restricted MC) 
derivatives of $\Psi$ with respect to $\epsilon$ are
still approximately equal at $T \approx T_S$~\cite{SM}. Then,
$\chi_s\approx{c(\tilde\lambda)/{\tilde g}}=\frac{\tilde g}{a_0(T_S-T^*)}$,
leading to the novel result
\vskip -0.3cm
\begin{equation}
T_S=T^*+{\tilde g^2\over{a_0c_0(1-{\tilde\lambda^2\over{c_0e_0}})}}.
\label{tslgnum}
\end{equation}

\begin{figure}[thbp]
\begin{center}
\includegraphics[trim = 1mm 0mm -1mm 0mm,width=0.5\textwidth,angle=0]{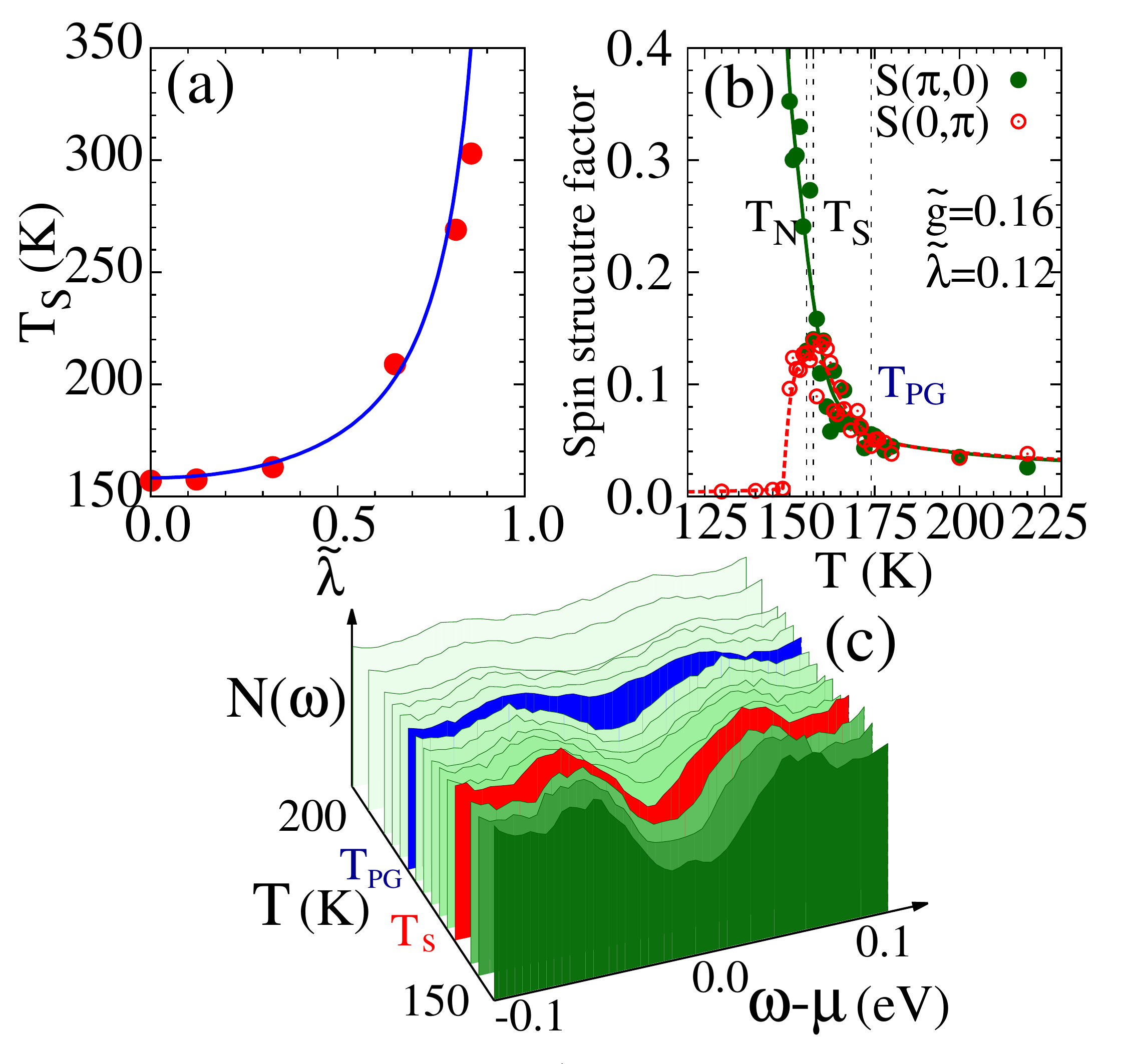}\label{fig:5a}
\vskip -0.40cm
\caption{(color online)
(a) The MC structural transition temperature $T_S$ vs. the orbital-lattice
coupling $\tilde \lambda$, at fixed $\tilde g = 0.16$. The continuous line is the fit
in Eq.(\ref{tslgnum}), from the GL equations.
(b) Spin structure factor $S({\bf k})$ vs. $T$ for the
two magnetic wavevectors of relevance. Results were obtained via MC
simulations on PBC 8$\times$8 clusters.
$T_{PG}$ is the pseudogap temperature [Fig.~\ref{fig:5}(c)].
(c) Density of states $N(\omega)$ (symmetrized) from unrestricted 
MC simulations on 8$\times$8 clusters ($\tilde g$=$0.16$; 
$\tilde \lambda$=$0.12$), at various $T$s. 
Results at $T_S$=$158$~K are 
in red. $T_{PG}$$\sim$$174$~K (blue) is the crossover $T$
where the pseudogap opens at the Fermi level ($\omega$-$\mu$=$0.0$) upon cooling.
}
\vskip -0.4cm
\label{fig:5}
\end{center}
\end{figure}

\noindent Numerically, it was found 
that $a_0$$\sim$$0.093$, $c_0$$\sim$$60$
$e_0$=$0.015$, and $T^*$=$105$~K, for $\tilde g$=$0.16$.
In practice, it was observed that Eq.(\ref{tslgnum}) fits remarkably well
the numerical values for $T_S$ in the $\tilde\lambda$-range studied showing
that the GL approach provides an excellent rationalization of the numerical results.
This is shown explicitly in Fig.~\ref{fig:5}(a).
%


{\it Spin structure factors and pseudogaps.}
In Fig.~\ref{fig:5}(b), 
the MC-calculated spin structure factor $S({\bf k})$ 
at both $(\pi,0)$ and $(0,\pi)$ are shown. 
The results illustrate the development of 
short-range magnetic order upon cooling with 
two coexisting wavevectors.
Within the error bars, given roughly by the oscillations in the plot, these results
indicate that the two wavevectors develop with equal
weight upon cooling approximately starting at $T_{PG}$ 
where the pseudogap develops (see below)~\cite{note}.

In the spin-fermion model, dynamical observables can
be easily calculated. 
In particular, the density of states $N(\omega)$ is shown in 
Fig.~\ref{fig:5}(c). This figure indicates the presence of a Fermi-level 
pseudogap (PG) in a wide temperature range, as in 
photoemission and infrared experiments~\cite{PG-exper}. 
A zero temperature pseudogap is to be expected: Hartree-Fock
studies of the
multiorbital Hubbard model~\cite{rong} already detected such a feature. 
However, our finite temperature studies reveal that upon cooling this pseudogap
develops at a $T_{PG}$ clearly above $T_S$. The pseudogap 
is present when short-range spin correlations 
are present [Fig.~\ref{fig:5}(b)]: 
the ``nematic fluctuations'' regime is basically
the $T$-range where $(\pi,0)$/$(0,\pi)$ magnetic fluctuations
exist. The coupling to the 
lattice creates concomitant local orthorrombic distortions: the region between
$T_S$ and $T_{PG}$ is tetragonal only on average~\cite{egami}.
All these results are in good agreement with recent scanning tunneling
spectroscopy studies of NaFeAs~\cite{rosenthal}.

{\it Conclusions.}
Our combined numerical and analytical study of the spin fermion model
leads to results in agreement with 
the experimentally measured nematic susceptibility 
of Ba(Fe$_{1-x}$Co$_x$)$_2$As$_2$~\cite{fisher-science}. 
In our analysis, which was time consuming and required a new MC setup,
magnetism is the main driver, but the lattice/orbital
are crucial to boost critical temperatures. For spins coupled to
the lattice our spin-nematic susceptibility 
has a Curie-Weiss behavior for $T>T_S$
governed by a $T^*$ which we here identify as the 
critical $T_N$ of the purely electronic sector, which is preexisting to 
the introduction of the lattice. 
For realistic nonzero electron-lattice 
couplings, the lattice induces a nematic/structural transition 
at a higher temperature $T_S$. 
The addition of an orbital-lattice coupling $\tilde \lambda$ further increases
$T_S$, although the Curie-Weiss behavior continues being regulated by $T^*$. 
Our main prediction is that whenever fluctuating nematic order 
is observed, inelastic neutron scattering 
for the same sample should also reveal the
existence of short-range magnetic order: nematic fluctuations, pseudogap, 
and short-range antiferromagnetic order should all develop
{\it simultaneously} in these materials. 

{\it Acknowledgments.} Conversations with Weicheng Lv are acknowledged.
S.L., A.M., and N.P. were supported by the 
National Science Foundation Grant No. DMR-1104386.
E.D. and A.M. were 
supported by the U.S. Department of Energy, Office of Basic Energy Sciences, 
Materials Sciences and Engineering Division.

\clearpage
\renewcommand{\thesection}{S\arabic{section}}
\renewcommand{\thesubsection}{S\arabic{subsection}}
\renewcommand{\theequation}{S\arabic{equation}}
\renewcommand{\thefigure}{S\arabic{figure}}
\setcounter{figure}{0}
\setcounter{equation}{0}

\section{Supplementary Material}

This Supplementary Material provides additional detail about
results presented in the main text. In particular, it includes: 
the full Hamiltonian, the derivations of equations deduced in
the Ginzburg-Landau context, and
Monte Carlo results at the (unphysically large~\cite{PRL-2013-SM}) coupling
$\tilde \lambda = 0.84$.

\subsection{Full Hamiltonian}

The full Hamiltonian of the 
spin-fermion model with lattice interactions incorporated is here provided. 
The same Hamiltonian was also used in Ref.~\cite{PRL-2013-SM}.
The model is given by:
\vspace{-0.2cm}
\begin{equation}
H_{\rm SF} = H_{\rm Hopp} + H_{\rm Hund} + H_{\rm Heis} + H_{\rm SL} + H_{\rm OL} + H_{\rm Stiff}.
\label{hamap}
\end{equation}
\vspace{-0.2cm}
\noindent The hopping component is made of three contributions,
\begin{equation}
H_{\rm Hopp}=H_{xz,yz}+H_{xy}+H_{xz,yz;xy}. 
\label{hop}
\end{equation}
\vspace{-0.2cm}
\noindent The first  term involves the $xz$ and $yz$ orbitals:
\begin{equation}\begin{split}
H_{xz,yz}&=
\{-t_1\sum_{{\bf i},\sigma}(d^{\dagger}_{{\bf i},xz,\sigma}
d^{\phantom{\dagger}}_{{\bf i}+\hat y,xz,\sigma}+d^{\dagger}_{{\bf i},yz,\sigma}
d^{\phantom{\dagger}}_{{\bf i}+\hat x,yz,\sigma}) \\
&-t_2\sum_{{\bf i},\sigma}(d^{\dagger}_{{\bf i},xz,\sigma}
d^{\phantom{\dagger}}_{{\bf i}+\hat x,xz,\sigma}+d^{\dagger}_{{\bf i},yz,\sigma}
d^{\phantom{\dagger}}_{{\bf i}+\hat y,yz,\sigma}) \\
&-t_3\sum_{{\bf i},\hat\mu\not=\hat\nu,\sigma}(d^{\dagger}_{{\bf i},xz,\sigma}
d^{\phantom{\dagger}}_{{\bf i}+\hat\mu+\hat\nu,xz,\sigma}+d^{\dagger}_{{\bf i},yz,\sigma}
d^{\phantom{\dagger}}_{{\bf i}+\hat\mu+\hat\nu,yz,\sigma}) \\
&+t_4\sum_{{\bf i},\sigma}(d^{\dagger}_{{\bf i},xz,\sigma}
d^{\phantom{\dagger}}_{{\bf i}+\hat x+\hat y,yz,\sigma}+d^{\dagger}_{{\bf i},yz,\sigma}
d^{\phantom{\dagger}}_{{\bf i}+\hat x+\hat y,xz,\sigma}) \\
&-t_4\sum_{{\bf i},\sigma}(d^{\dagger}_{{\bf i},xz,\sigma}
d^{\phantom{\dagger}}_{{\bf i}+\hat x-\hat y,yz,\sigma}+d^{\dagger}_{{\bf i},yz,\sigma}
d^{\phantom{\dagger}}_{{\bf i}+\hat x-\hat y,xz,\sigma})\\
&+h.c.\}-\mu\sum_{\bf i}(n_{{\bf i},xz}+n_{{\bf i},yz}).
\label{ham12}
\end{split}\end{equation}
\noindent The second term contains the hoppings related with the $xy$ orbital:
\begin{equation}\begin{split}
H_{xy}=&\ t_5\sum_{{\bf i},{\hat \mu},\sigma}(d^{\dagger}_{{\bf i},xy,\sigma}
d^{\phantom{\dagger}}_{{\bf i}+\hat \mu,xy,\sigma}+h.c.)\\
&-t_6\sum_{{\bf i},\hat\mu\not=\hat\nu,\sigma}(d^{\dagger}_{{\bf i},xy,\sigma}
d^{\phantom{\dagger}}_{{\bf i}+\hat\mu+\hat\nu,xy,\sigma}+h.c.)\\
&+\Delta_{xy}\sum_{\bf i}n_{{\bf i},xy}-\mu\sum_{\bf i}n_{{\bf i},xy},\;
\label{ham3}
\end{split}\end{equation}
\noindent The last hopping term is:
\begin{equation}\begin{split}
H_{\rm xz,yz;xy}=&-t_7\sum_{{\bf i},\sigma}[(-1)^{|{\bf i}|}d^{\dagger}_{{\bf i},xz,\sigma}
d^{\phantom{\dagger}}_{{\bf i}+\hat x,xy,\sigma}+h.c.]\\
&-t_7\sum_{{\bf i},\sigma}[(-1)^{|{\bf i}|}d^{\dagger}_{{\bf i},xy,\sigma}
d^{\phantom{\dagger}}_{{\bf i}+\hat x,xz,\sigma}+h.c.]\\
&-t_7\sum_{{\bf i},\sigma}[(-1)^{|{\bf i}|}d^{\dagger}_{{\bf i},yz,\sigma}
d^{\phantom{\dagger}}_{{\bf i}+\hat y,xy,\sigma}+h.c.] \\
&-t_7\sum_{{\bf i},\sigma}[(-1)^{|{\bf i}|}d^{\dagger}_{{\bf i},xy,\sigma}
d^{\phantom{\dagger}}_{{\bf i}+\hat y,yz,\sigma}+h.c.] \\
&-t_8\sum_{{\bf i},\sigma}[(-1)^{|{\bf i}|}d^{\dagger}_{{\bf i},xz,\sigma}
d^{\phantom{\dagger}}_{{\bf i}+\hat x+\hat y,xy,\sigma}+h.c.] \\
&+t_8\sum_{{\bf i},\sigma}[(-1)^{|{\bf i}|}d^{\dagger}_{{\bf i},xy,\sigma}
d^{\phantom{\dagger}}_{{\bf i}+\hat x+\hat y,xz,\sigma}+h.c.] \\
&-t_8\sum_{{\bf i},\sigma}[(-1)^{|{\bf i}|}d^{\dagger}_{{\bf i},xz,\sigma}
d^{\phantom{\dagger}}_{{\bf i}+\hat x-\hat y,xy,\sigma}+h.c.]\\
&+t_8\sum_{{\bf i},\sigma}[(-1)^{|{\bf i}|}d^{\dagger}_{{\bf i},xy,\sigma}
d^{\phantom{\dagger}}_{{\bf i}+\hat x-\hat y,xz,y\sigma}+h.c.] \\
&-t_8\sum_{{\bf i},\sigma}[(-1)^{|{\bf i}|}d^{\dagger}_{{\bf i},yz,\sigma}
d^{\phantom{\dagger}}_{{\bf i}+\hat x+\hat y,xy,\sigma}+h.c.] \\
&+t_8\sum_{{\bf i},\sigma}[(-1)^{|{\bf i}|}d^{\dagger}_{{\bf i},xy,\sigma}
d^{\phantom{\dagger}}_{{\bf i}+\hat x+\hat y,yz,\sigma}+h.c.] \\
&+t_8\sum_{{\bf i},\sigma}[(-1)^{|{\bf i}|}d^{\dagger}_{{\bf i},yz,\sigma}
d^{\phantom{\dagger}}_{{\bf i}+\hat x-\hat y,xy,\sigma}+h.c.] \\
&-t_8\sum_{{\bf i},\sigma}[(-1)^{|{\bf i}|}d^{\dagger}_{{\bf i},xy,\sigma}
d^{\phantom{\dagger}}_{{\bf i}+\hat x-\hat y,yz,\sigma}+h.c.].
\label{ham12_3}
\end{split}\end{equation}
In the equations above, 
the operator $d^{\dagger}_{{\bf i},\alpha,\sigma}$ creates an electron at 
site ${\bf i}$ of the two-dimensional lattice of irons. The orbital index is $\alpha=$  
$xz$, $yz$, or $xy$, and the $z$-axis spin projection is $\sigma$. The chemical potential
used to regulate the electronic density is $\mu$. 
The symbols ${\hat x}$ and ${\hat y}$ denote
vectors along the axes that join NN atoms.
The values of the hoppings $t_i$ are from Ref.~\cite{three-SM} and 
they are reproduced in Table~\ref{tableap}, 
including also the value of the energy splitting $\Delta_{xy}$.
\begin{table}
\caption{Values of the parameters that appear in the tight-binding portion of the three-orbital model
  Eqs.(\ref{ham12}) to (\ref{ham12_3}). The overall energy unit is
  electron volts.\label{tableap}}
 \begin{tabular}{|ccccccccc|}\hline
$t_1$ & $t_2$ & $t_3$ & $t_4$ & $t_5$ & $t_6$ & $t_7$ & $t_8$ &
   $\Delta_{xy}$\\
\hline
  0.02   &0.06    &0.03   &$-0.01$&$0.2$ & 0.3 & $-0.2$ & $0.1$&
  0.4\\ \hline
 \end{tabular}
\end{table}

The remaining terms of the Hamiltonian have been briefly discussed in the main text.
The symbols $\langle \rangle$ denote NN while $\langle\langle \rangle\rangle$ denote NNN. 
The rest of the notation is standard.  
\begin{equation}
H_{\rm Hund}= -{J_{\rm H}}\sum_{{\bf i},\alpha} {{{\bf S}_{\bf i}}\cdot{{\bf s}_{{\bf i},\alpha}}}, 
\label{hund}
\end{equation}
\vspace{-0.2cm}
\begin{equation}
H_{\rm Heis}= J_{{\rm NN}}\sum_{\langle{\bf ij}\rangle} {\bf S}_{{\bf i}}\cdot{\bf S}_{{\bf j}}
+J_{{\rm NNN}}\sum_{\langle\langle{\bf im}\rangle\rangle} {\bf S}_{{\bf i}}\cdot{\bf S}_{{\bf m}}, 
\label{heis}
\end{equation}
\vspace{-0.2cm}
\begin{equation}
H_{\rm SL}=-g\sum_{\bf i}\Psi_{\bf i}\epsilon_{\bf i}, 
\label{sl}
\end{equation}
\vspace{-0.2cm}
\begin{equation}
H_{\rm OL}=-\lambda\sum_{\bf i}\Phi_{\bf i}\epsilon_{\bf i}, 
\label{ol}
\end{equation}
\vspace{-0.2cm}
\begin{equation}\begin{split}
H_{\rm Stiff}= {1\over{2}}k\sum_{\bf i}\sum_{\nu=1}^4(|{\bf R}^{\bf i \nu}_{Fe-As}|-R_0)^2+\\
+k'\sum_{<{\bf ij}>}[({a_0\over{R^{\bf ij}_{Fe-Fe}}})^{12}-
2({a_0\over{R^{\bf ij}_{Fe-Fe}}})^6].
\label{stiff}
\end{split}
\end{equation}
\vspace{-0.2cm}


The $\mathcal{O}_{rth}$ strain $\epsilon_{\bf i}$ is defined as:

\vspace{-0.2cm}
\begin{equation}
\epsilon_{\bf i}= {1\over{4\sqrt{2}}}\sum_{\nu=1}^4 (|\delta_{\bf i,\nu}^y|-|\delta_{\bf i,\nu}^x|), 
\vspace{-0.1cm}
\label{epsilon}
\end{equation}
where $\delta_{\bf i,\nu}^x$($\delta_{\bf i,\nu}^y$) 
is the component along $x$ ($y$) of the distance 
between the Fe atom at site ${\bf i}$ of the lattice 
and one of its four neighboring  As atoms that are labeled
 by the index $\nu=1, 2, 3, 4$. For more details of the notation used 
see Ref.~\cite{PRL-2013-SM}, where the technical aspects on how to 
simulate an orthorrombic distortion 
can also be found.

%
%

\subsection{Ginzburg-Landau phenomenological approach}

In this section, the Monte Carlo data gathered for the spin-fermion model
will be described via a phenomenological Ginzburg-Landau (GL) approach, to provide
a more qualitative description of those numerical results. More specifically, 
the free energy $F$ of the SF model will be (approximately) written 
in terms of the spin-nematic order parameter $\Psi$, 
the orbital-nematic order parameter 
$\Phi$, and the orthorhombic strain $\epsilon$, as in GL descriptions. 
In previous literature a single nematic order parameter was considered
without separating its magnetic 
and orbital character~\cite{fisher-science-SM,kasahara-SM,fernandes2-SM}. 
In addition, it was necessary
to formulate assumptions about the order of the nematic 
and structural transitions. 
In our case, the MC results in this and previous publications 
are used as guidance to address this matter at the free energy level.
More specifically, a second order magnetic transition was previously
reported for the purely electronic system~\cite{PRL-2012-SM}. Thus, 
the spin-nematic portion of $F$ should display a free energy with 
a second order phase transition. 

With regards to the terms involving $\epsilon$, the MC results 
of Ref.~\cite{PRL-2013-SM}
showed that the coupling of the spin-nematic order parameter to the lattice 
leads to a weak first order (or very sharp second order) nematic and structural 
transition.  Naively, this implies that the order $\epsilon^4$ term should have a
negative coefficient. However, since in our numerical simulations 
a Lennard-Jones potential is used for the elastic term, then the sign of the quartic term is fixed and it happens to
be positive. However, considering that the $\epsilon$ 
displacements are very small and the transition is weakly first order at best, 
then just the harmonic (second order) approximation 
should be sufficient for $\epsilon$. 

After all these considerations, the free energy is given by:
\begin{equation}
F={a\over{2}}\Psi^2+{b\over{4}}\Psi^4+{c\over{2}}\epsilon^2+{e\over{2}}\Phi^2+{f\over{4}}\Phi^4
\end{equation}
\begin{equation}
-\tilde g \Psi\epsilon-\tilde\lambda\Phi\epsilon-h\epsilon,
\label{F}
\end{equation}
\noindent where $a$, $b$, $c$, $e$, and $f$ are the coefficients 
of the many terms of the three order parameters, while $\tilde g$ 
and $\tilde\lambda$ are the coupling constants of the lattice with the
spin and orbital degrees of freedom as described in the main text.
Since this and previous MC studies~\cite{PRL-2012-SM,PRL-2013-SM} 
showed that there is no long-range 
orbital order in the ground state of the SF model, at least in the
range of couplings investigated,
then a positive quartic term is used for this order parameter.
The parameter $h$ denotes an external stress, 
as explained in Ref.~\cite{fisher-science-SM}. 
Note that in principle another term, and associated 
coupling constant, ${\tilde \alpha}\Psi \Phi$ should be included in $F$. This
term will affect the orbital susceptibility deduced at the end of this subsection. However, adding this term requires varying another parameter 
in the SF model MC simulation, thus increasing substantially
the time demands for this project. As a consequence, this addition 
is postponed for the near future.

As explained in the main text,
our MC results indicate that the leading order parameter guiding the
results is the spin-nematic $\Psi$. Thus, it is reasonable to assume that 
only the coefficient $a$ depends on temperature as
$a=a_0(T-T^*)$, while other parameters, such as  
$c=c_0$ (the uncoupled shear elastic modulus) 
and $e=e_0$, are approximately $T$-independent.

For the special case $\tilde g=\tilde\lambda=0$ 
the critical temperature $T^*$ for the magnetic transition 
can be obtained by setting to zero the derivative of $F$ with respect to $\Psi$:
\begin{equation}
{\partial F\over{\partial\Psi}}=a\Psi+b\Psi^3=0.
\label{Fpsi00}
\end{equation}
Then, for $T\le T^*$ the order parameter is given by
\begin{equation}
\Psi=\sqrt{{a_0\over{b}}(T^*-T)}.
\label{PsiT}
\end{equation}
\noindent The equation above is valid only when $\Psi$ is small, 
i.e. close to the transition temperature from below. Additional terms in the free energy 
would be needed as $T\rightarrow 0$ since in that limit $|\Psi|=2$. 

Now consider the case when $\tilde g$ is nonzero, still keeping $\tilde\lambda=0$.
Setting to zero the derivative of $F$ with respect to $\Psi$ and $\epsilon$ 
leads to (for $h=0$):
\begin{equation}
{\partial F\over{\partial\epsilon}}=c_0\epsilon-\tilde g \Psi=0,
\label{Feps0}
\end{equation}
\begin{equation}
{\partial F\over{\partial\Psi}}=a\Psi+b\Psi^3-\tilde g \epsilon=0.
\label{Fpsi0}
\end{equation}
From Eq.(\ref{Feps0}),
\begin{equation}
\Psi={c_0\over{\tilde g}}\epsilon,
\label{linear}
\end{equation}
\noindent 
which reproduces the linear relation obtained 
numerically before, see Fig.~\ref{newfig}(b) main text, with a slope now explicitly 
given in terms of $\tilde g$ and a constant that 
now can be identified with the bare shear elastic modulus $c_0$.

Solving for $\epsilon$ in Eq.(\ref{Fpsi0}) and introducing 
the result in Eq.(\ref{Feps0}) leads to:
\begin{equation}
(a-{\tilde g^2\over{c_0}})\Psi+b\Psi^3=0,
\label{Feps0rep}
\end{equation}
\noindent where it is clear that $a$ becomes renormalized 
due to the coupling to the lattice. The transition now occurs 
at a renormalized temperature $T_S$ that satisfies:
\begin{equation}
a_0(T-T_S)=a-{\tilde g^2\over{c_0}}=a_0(T-T^*)-{\tilde g^2\over{c_0}}.
\label{tts}
\end{equation}
\noindent From the expression above, 
it can be shown that the new nematic transition occurs at
\begin{equation}
T_S=T^*+{\tilde g^2\over{a_0c_0}},
\label{tsa}
\end{equation}
\noindent and clearly $T_S>T^*$. Note that Eq.(\ref{tsa}) 
has been obtained in previous GL analysis, but in those
studies a generic nematic coupling appeared in the numerator of the second
term while here, more specifically, 
we identify $\tilde g$ with the spin-nematic coupling to the lattice.

Reciprocally, solving for $\Psi$ in Eq.(\ref{Feps0}) and introducing the result in Eq.(\ref{Fpsi0})
leads to:
\begin{equation}
{a\over{\tilde g}}[(c_0-{\tilde g^2\over{a}})\epsilon+{bc_0^3\over{\tilde g^2a}}\epsilon^3]=0,
\label{Fpsi0rep}
\end{equation}
\noindent where, due to the coupling to the lattice, now 
the shear constant is renormalized and an effective quartic term 
is generated for the lattice free energy. The effective shear elastic 
modulus $c_{66}$ becomes temperature dependent and it is given by:
\begin{equation}
c_{66}=c_0-{\tilde g^2\over{a_0(T-T^*)}},
\label{c66}
\end{equation}
\noindent that vanishes at $T=T_S$. Thus, the structural transition 
occurs at the same critical temperature $T_S$ of the nematic transition.

To obtain the spin-nematic susceptibility, the
second derivative of $F$ with respect to $\Psi$ and $h$ is set to zero:
\begin{equation}
{\partial^2 F\over{\partial h\partial\Psi}}=a{\partial\Psi\over{\partial h}}+
3b\Psi^2{\partial\Psi\over{\partial h}}-\tilde g {\partial\epsilon\over{\partial h}}=0,
\label{secondder}
\end{equation}
\noindent and then
\begin{equation}
\chi_s={\partial\Psi\over{\partial\epsilon}}={{\partial\Psi\over{\partial h}}\over{{\partial\epsilon\over{\partial h}}}}=
{\tilde g\over{a+3b\Psi^2}}={\tilde g\over{a_0(T-T^*)+3b\Psi^2}}.
\label{chisapl0}
\end{equation}
\noindent This is an important equation that was used in the main text to rationalize
the MC numerical results.
In the range $T\ge T_S$, i.e. when $\Psi=0$, the spin-nematic susceptibility 
clearly follows a Curie-Weiss behavior. In practice, it has been observed 
that $b=a_0T_S$ to a good approximation.

Consider now the case when the orbital-lattice 
coupling $\tilde\lambda$ is nonzero as well. Now
\begin{equation}
{\partial F\over{\partial\epsilon}}=c_0\epsilon-\tilde g \Psi-\tilde\lambda\Phi=0,
\label{Fepsl}
\end{equation}
\begin{equation}
{\partial F\over{\partial\Psi}}=a\Psi+b\Psi^3-\tilde g \epsilon=0,
\label{Fpsi0l}
\end{equation}
\noindent and a new equation is available:
\begin{equation}
{\partial F\over{\partial\Phi}}=e_0\Phi+f\Phi^3-\tilde\lambda\epsilon=0.
\label{Fphil}
\end{equation}
Solving for $\Psi$ in Eq.(\ref{Fepsl}) leads to:
\begin{equation}
\Psi={c_0\epsilon-\tilde\lambda\Phi\over{\tilde g}},
\label{psil}
\end{equation}
\noindent while solving for $\epsilon$ in Eq.(\ref{Fpsi0l}) leads to:
\begin{equation}
\epsilon={a\Psi+b\Psi^3\over{\tilde g}}.
\label{epsil}
\end{equation}
Introducing Eq.(\ref{epsil}) into Eq.(\ref{psil}), $\Phi$ is obtained in terms of $\Psi$ as follows:
\begin{equation}
\Phi=({c_0\over{\tilde\lambda\tilde g}})[(a-{\tilde g^2\over{c_0}})\Psi+b\Psi^3].
\label{phipsi}
\end{equation}
Introducing Eqs.(\ref{epsil}) and (\ref{phipsi}) into Eq.(\ref{Fphil}) 
a renormalized equation for $\Psi$ is obtained:
\begin{equation}
[{e_0c_0\over{\tilde\lambda\tilde g}}(a-{\tilde g^2\over{c_0}})-{\tilde\lambda a\over{\tilde g}}]\Psi\
+[{e_0c_0\over{\tilde\lambda\tilde g}}b-
{\tilde\lambda b\over{\tilde g}}+{fc_0^3\over{\tilde\lambda^3\tilde g^3}}(a-{\tilde g^2\over{c_0}})^3\
]\Psi^3=0.
\label{Fphiln}
\end{equation}
Then, at $T=T_S$ the effective coefficient of the linear term in $\Psi$ 
provides the new transition temperature:
\begin{equation}
a_0(T-T_S)=a-{e_0\tilde g^2\over{e_0c_0-\tilde\lambda^2}}.
\label{tsl}
\end{equation}
\noindent 
Using that $a=a_0(T-T^*)$, the dependence of the
critical temperature with the two coupling constants $\tilde g$ and $\tilde \lambda$ can be obtained:
\begin{equation}
T_S=T^*+{\tilde g^2\over{a_0c_0(1-{\tilde\lambda^2\over{c_0e_0}})}}.
\label{tslg}
\end{equation}
\noindent This is another interesting 
formula that nicely describes the MC results, as shown in the
main text. Equation(\ref{tslg}) is a novel result that 
shows that $T_S$ depends in a different way on
the spin-lattice ($\tilde g$) and 
the orbital-lattice ($\tilde\lambda$) couplings.
Moreover, an effective $\tilde\lambda$-dependent  
elastic modulus $c(\tilde\lambda)$ can be defined as
\begin{equation}
c(\tilde\lambda)=c_0-{\tilde\lambda^2\over{e_0}}.
\label{ceff}
\end{equation}
In addition, the effective shear elastic modulus is now given by
\begin{equation}
c_{66}=c_0-{\tilde\lambda^2\over{e_0}}-{\tilde g^2\over{a_0(T-T^*)}},
\label{c66n}
\end{equation}
\noindent which vanishes at the $T_S$ given by Eq.(\ref{tslg}).

The spin-nematic susceptibility is still given by Eq.(\ref{chisapl0}) 
with the dependence on $\tilde\lambda$ embedded in the actual values of $\Psi$.
The orbital-nematic susceptibility is obtained from Eq.(\ref{Fphil}) as
\begin{equation}
{\partial^2 F\over{\partial h\partial\Phi}}=(e_0+3f\Phi^2){\partial\Phi\over{\partial h}}-\tilde\lambda\
 {\partial\epsilon\over{\partial h}}-\tilde\alpha {\partial\Phi\over{\partial h}}=0.
\label{double2}
\end{equation}
In the absence of an explicit coupling $\tilde\alpha$ between 
the spin-nematic and orbital order parameters, then 
the orbital-nematic susceptibility becomes:
\begin{equation}
\chi_o={\partial\Phi\over{\partial\epsilon}}={{\partial\Phi\over{\partial h}}\over{{\partial\epsilon\over{\partial h}}}}={\tilde\lambda\over{e_0+3f\Phi^2}}.
\label{chioapa0}
\end{equation}


\subsection{Partial and total derivatives at $T_S$}

The partial derivative in the definition of $\chi_s$ is at constant $T$ varying
$\epsilon$ and it is evaluated at equilibrium $\epsilon=\epsilon_0$. The slopes
of the green and blue curves of Fig.~\ref{newfig}(b) in the main text 
provide this derivative. On the
other hand, the results of Fig.~\ref{newfig}(b) in equilibrium (slope of the 
red points curve) provide the full derivative ${d\Psi\over{d\epsilon}}$. 
Since $\epsilon$=$\epsilon(T)$, their relation is
\vskip -0.1cm
\begin{equation}
{d\Psi\over{d\epsilon}}={\partial\Psi\over{\partial\epsilon}}|_{\epsilon_0}
+{\partial\Psi\over{\partial T}}|_{\epsilon_0}{\partial T\over{\partial\epsilon}}|_{\epsilon_0}=
\chi_s+{{\partial\Psi\over{\partial T}}|_{\epsilon_0}\over{{\partial\epsilon\over{\partial T}}|_{\epsilon_0}}},
\label{derivatives}
\end{equation}
\vskip -0.1cm
\noindent where ${\partial\Psi\over{\partial T}}$ 
is performed at constant $\epsilon$ and ${\partial\epsilon\over{\partial T}}|_{\epsilon_0}$ 
is performed at constant $\Psi$. In general, the partial and total 
derivatives of $\Psi$ with respect to $\epsilon$ can differ from one another.
However, at small $\tilde\lambda$ 
the structural transition is weakly first order~\cite{PRL-2013-SM} 
(or a very sharp second order) and then when $T \approx T_S$
the lattice distortion $\epsilon$ rapidly jumps from 0 to a finite value.
This means that ${\partial\epsilon\over{\partial T}}|_{\epsilon_0}$ 
is very large while
${\partial\Psi\over{\partial T}}|_{\epsilon_0}$ remains finite since it is 
performed at fix $\epsilon$. Thus, at $T \approx T_S$, the partial and total 
derivatives are almost the same. This can be seen in Fig.~\ref{newfig}(b) of
the main text where the slopes of
the green curves at $\epsilon=0$, when they 
cross the equilibrium line, are smaller than the 
equilibrium slope $K$ but increase with decreasing $T$ until it becomes equal 
to $K$ at $T=T_S$ (red line). The slopes of the blue curves 
at the finite value of $\epsilon$ where they cross the equilibrium 
line are smaller than $K$ and decrease with decreasing $T$. 



\subsection{Spin-nematic susceptibility at large $\tilde \lambda$}

\begin{figure}[thbp]
\includegraphics[trim = 7mm 0mm -7mm 0mm,width=0.57\textwidth,angle=0]{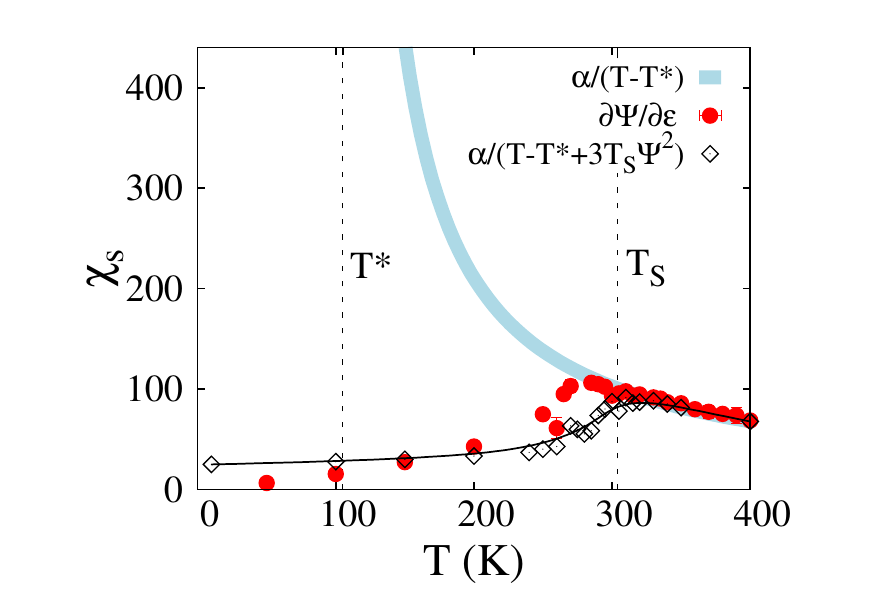}\label{sup:1a}
\vskip -0.3cm
\caption{(color online) 
Spin-nematic susceptibility $\chi_s$ vs. temperature $T$ (red circles) obtained 
from Fig.~\ref{newfig.SM}(b) (at $\tilde g$=$0.16$ and $\tilde \lambda$=$0.84$). 
The standard MC technique on an 8$\times$8 cluster with PBC
was employed (involving ED of the fermions at every MC step).
Also shown are two GL fits, as also employed in Fig.~\ref{fig:1}. 
The blue (thick) line 
indicates a divergence at a temperature $T^*$ (lower than $T_S$) characteristic 
of the electronic sector alone. In the range $T \leq T_S$, the lattice follows 
the electronic behavior. The black (thin) line and black tilted square points 
are a fit including the $3T_S\Psi^2$ correction (see text in the previous 
section of this Suppl. Material). The fitting parameters 
are $T^* = 105$~K and $T_S=304$~K. The actual 
N\'eel temperature for $\tilde g$=$0.16$ 
and $\tilde \lambda$=$0.84$ is not shown. 
}
\label{sup:1}
\end{figure}

To investigate in more detail the potential role of orbital
order in the spin-nematic susceptibility, simulations were repeated
for a robust $\tilde \lambda=0.84$, keeping the other electron-lattice 
coupling fixed as $\tilde g =0.16$. 
Results are shown in Fig.~\ref{sup:1}.
The increase of $\tilde \lambda$ substantially increases $T_S$, which is to
be expected since now the electron-lattice coupling is larger~\cite{PRL-2013-SM}. 
However, above $T_S$ still the results can be well fit by a Curie-Weiss law,
with a divergence at $T^*$ which is the critical temperature of the purely
electronic system, as described in the main text. Even the coefficient
$a_0$ in the fit is almost identical 
to that of the case $\tilde \lambda=0.12$, in Fig.~\ref{fig:1}. 
The second fit, with the 3$T_S$$\Psi^2$ correction, is still reasonable.
In summary, as long as $\tilde \lambda$ is not increased to such large values that
the low-temperature ground state is drastically altered, the computational results 
can still be analyzed via the GL formalism outlined here and in the main text, with
a $T^*$ that originates in the $(\pi,0)$ magnetic transition of the purely electronic
sector.

\begin{figure}[thbp]
\begin{center}
\includegraphics[trim = 5mm 0mm -5mm 0mm,width=0.44\textwidth,angle=0]{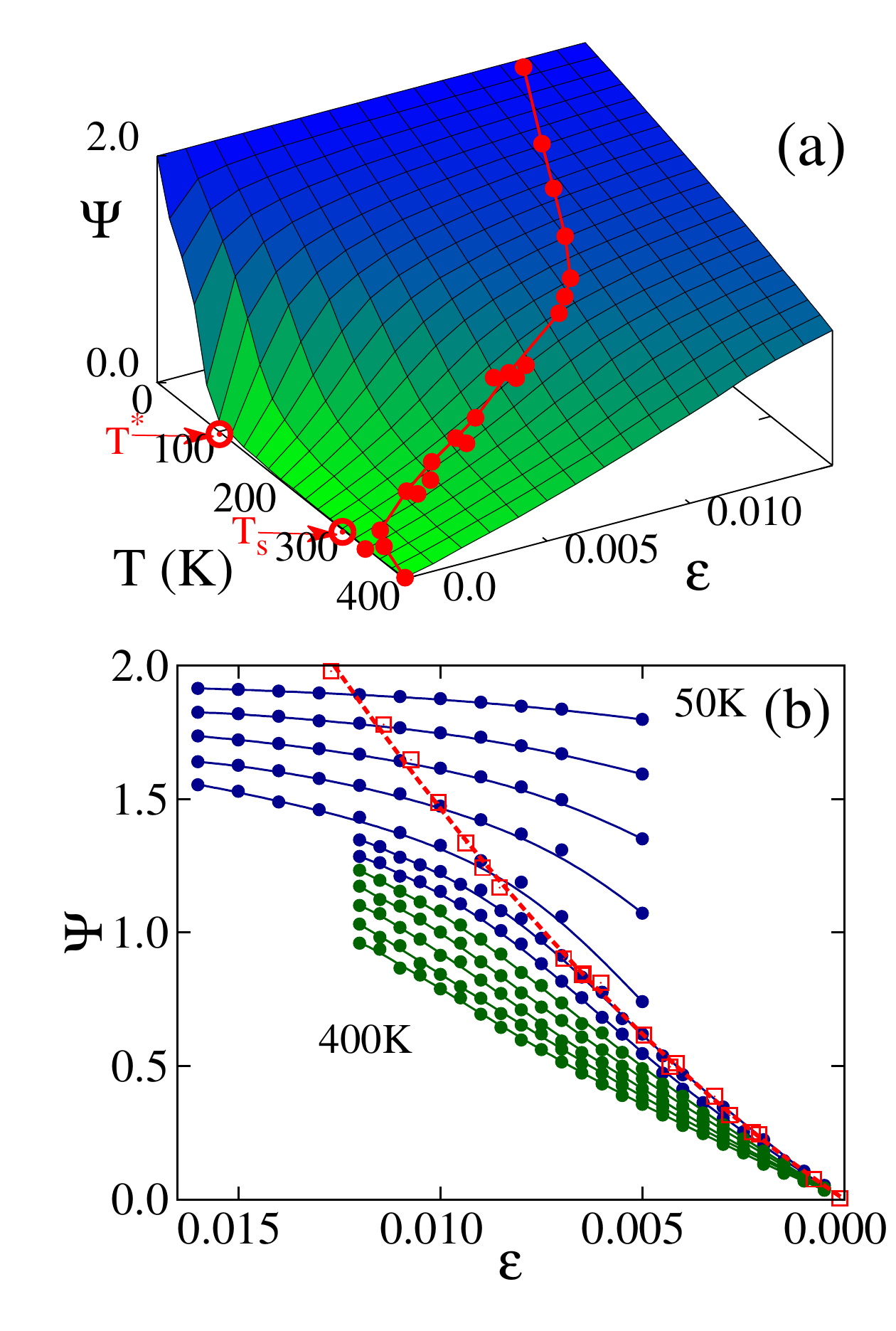}\label{newfig.SMa}
\vskip -0.3cm
\caption{(color online)
Spin-nematic order parameter from the MC simulations, 
at $\tilde g$=$0.16$ and $\tilde \lambda$=$0.84$.
(a) $\Psi$ vs. $T$ and $\epsilon$, 
measured at a fixed lattice distortion 
$\epsilon$ for each $T$ (restricted MC). Shown are the $T^*$ temperature 
(see text) and $T_S$.
 Results shown are for an 8$\times$8 cluster with TCA+TBC, but 
PBC 8$\times$8 clusters with ED give similar results. Red points 
are the equilibrium values using unrestricted MC with ED and PBC 8$\times$8 clusters.
(b) MC results illustrating the 
relation between $\Psi$ and $\epsilon$ in unrestricted MC (red) and 
the restricted MC curves (green/blue), parametric with $T$.
Results are obtained with ED/PBC 8$\times$8 clusters. 
Note that $\Psi$ vs. $\epsilon$ (red squares) 
is no longer linear which is expected because Eq.(\ref{linear}) is valid 
only for $~\tilde\lambda=0$ (and approximately valid for small $\tilde\lambda$).
}
\label{newfig.SM}
\end{center}
\end{figure}

For completeness, the plots analog to those of Fig.~\ref{newfig} but in
the present case of $\tilde \lambda=0.84$ are provided in Fig.~\ref{newfig.SM}.

\begin{thebibliography}{10}


\bibitem{johnston} D. C. Johnston, Adv. Phys. {\bf 59}, 803 (2010).

\bibitem{dai} P. Dai, J.-P. Hu , and E. Dagotto,
Nat. Phys. {\bf 8}, 709 (2012).

\bibitem{fisher} J-H. Chu, J. G. Analytis, K. De Greve, P. L. McMahon, Z.
Islam, Y. Yamamoto, and I. R. Fisher, Science {\bf 329}, 824
(2010); See also I. R. Fisher, L. Degiorgi, and Z. X. Shen,
Rep. Prog. Phys. {\bf 74}, 124506 (2011).

\bibitem{fradkin} E. Fradkin {\it et al.}, 
Annu. Rev. Cond. Mat. Phys. {\bf 1}, 153 (2010).


\bibitem{spin3} R. M. Fernandes {\it et al.}, 
Phys. Rev. Lett. {\bf 105}, 157003 (2010).

\bibitem{spin1} C. Fang, H. Yao, W.-F. Tsai, J.P. Hu, and S. A. Kivelson,
Phys. Rev. B {\bf 77}, 224509 (2008).

\bibitem{spin2} C. Xu, M. M\"uller, and S. Sachdev,
Phys. Rev. B {\bf 78}, 020501(R) (2008).


\bibitem{fernandes1} R. M. Fernandes {\it et al.}, 
Phys. Rev. B {\bf 85}, 024534 (2012).

\bibitem{fernandes2} R. M. Fernandes, A. V. Chubukov, 
and J. Schmalian, Nature Phys. {\bf 10}, 97 (2014).



\bibitem{bnl2} C.-C. Lee, W.-G. Yin, and Wei Ku,
Phys. Rev. Lett. {\bf 103}, 267001 (2009).

\bibitem{orb1} C.-C. Chen {\it et al.}, 
Phys. Rev. B {\bf 80}, 180418(R) (2009); C.-C. Chen {\it et al.}, 
Phys. Rev. B {\bf 82}, 100504(R) (2010).

\bibitem{orb2} W. Lv, J.S. Wu, and P. Phillips,
Phys. Rev. B {\bf 80}, 224506 (2009);
W.-C. Lee {\it et al.}, 
Phys. Rev. B {\bf 86}, 094516 (2012).


\bibitem{kontani} H. Kontani {\it et al.}, 
Solid State Comm. {\bf 152}, 718 (2012); H. Kontani, T. Saito, and S. Onari,
Phys. Rev. B {\bf 84}, 024528 (2011). 


\bibitem{PRL-2013} S. Liang, A. Moreo, 
and E. Dagotto, Phys. Rev. Lett. {\bf 111}, 047004 (2013).



\bibitem{BNL} W.-G. Yin, C.-C. Lee, and W. Ku, Phys. Rev. Lett. {\bf 105}, 107004 (2010).

\bibitem{kruger} W. Lv, F. Kr\"uger, and P. Phillips,
Phys. Rev. B {\bf 82}, 045125 (2010).

\bibitem{PRL-2012} S. Liang, G. Alvarez, C. Sen, A. Moreo, and E. Dagotto, 
Phys. Rev. Lett. {\bf 109}, 047001 (2012). 


\bibitem{CMR} E. Dagotto, T. Hotta, and A. Moreo, Phys. Rep. {\bf 344}, 1
(2001).

\bibitem{loca1} H. Gretarsson {\it et al.}, 
Phys. Rev. B {\bf 84}, 100509(R) (2011).

\bibitem{loca2} F. Bondino {\it et al.}, 
Phys. Rev. Lett. {\bf 101}, 267001 (2008).


\bibitem{wide} $\Delta_{SN}$=$T_S$-$T_N$ can be regulated by the
electron-orbital coupling $\tilde \lambda$ leading to a $\Delta_{SN}$ 
in our model
larger than the small values reported for spin systems
[see Y. Kamiya, N. Kawashima, and C. D. Batista,
Phys. Rev. B {\bf 84}, 214429 (2011); 
A. L. Wysocki, K. D. Belashchenko, and V. P. Antropov,
Nat. Phys. {\bf 7}, 485 (2011)].




\bibitem{fisher-science} 
J-H. Chu, H-H. Kuo, J. G. Analytis, and I. R. Fisher, Science {\bf 337}, 710 (2012); H.~H. Kuo {\it et al.}, Phys. Rev. B {\bf 88}, 085113 (2013);
and references therein.

\bibitem{SM} See Supplemental Material at
http://link.aps.org/supplemental/xx.xxxx
for details of the GL calculations and results at $\tilde \lambda$=$0.84$.


\bibitem{three} M. Daghofer {\it et al.}, 
Phys. Rev. B {\bf 81}, 014511 (2010).

\bibitem{clari} The original definitions of $\Psi$ and
$\epsilon$ in~\cite{PRL-2013} have been multiplied by $-1$ so that $\Psi$ and $\epsilon$
are both positive here, as assumed in the GL analysis. 

\bibitem{only} The spin in $H_{\rm SL}$ will only 
be the localized spin for computational simplicity.

\bibitem{kumar} S. Kumar and P. Majumdar, Eur. Phys. J. B {\bf 50}, 571 (2006).

\bibitem{steps} In unrestricted MC employing
the ED method on 8$\times$8 clusters, typically
8,000 thermalization (Th) and up to 100,000 measurement (Ms) steps were used.
In restricted MC with ED and 8$\times$8 clusters, 
the numbers are 8,000 and 20,000 for Th and Ms steps. 
In restricted MC using TCA+TBC, 4,000 Th and 4,000 Ms steps 
were employed for a 16$\times$16 cluster with
a 4$\times$4 cluster for the MC updates, while for an 8$\times$8 (same MC update cluster) the numbers 
were 20,000 for Th and 20,000 for Ms steps.


\bibitem{salafranca} J. Salafranca, G. Alvarez, and E. Dagotto, Phys. Rev. B {\bf 80}, 155133 (2009).


\bibitem{T8T16}
The value of $T^*$ is not the same (but close)
for 8$\times$8 and 16$\times$16 lattices due 
to size effects. Then,  the fits for each lattice size
are carried out with the $T^*$ of each cluster.









\bibitem{kasahara} S. Kasahara {\it et al.}, 
Nature {\bf 486}, 382 (2012).


\bibitem{orbitalsusce} The orbital-based nematic susceptibility,
$\chi_o=\frac{\partial \Phi}{\partial \epsilon}|_{\epsilon_0}$,
was also numerically calculated varying the temperature (not shown). 
For small $\tilde \lambda$, 
such as $\tilde\lambda=0.12$, the result is approximately temperature 
independent and well fit by Eq.(S27) in~\cite{SM}, with $\epsilon_0 = 0.015$ 
and $f= 0.33$. In other words, 
the analog of Fig.~\ref{newfig}(b) but for the orbital-nematic
order parameter presents blue/green/red curves all with very similar slopes.
Then, in $\chi_o$ there is no Curie-Weiss behavior for $T \ge T_S$ and 
in our model the orbital DOF plays a secondary role.
This is also in agreement with angle-resolved photoemission experiments that
reported a different population of the $d_{xz}$ and $d_{yz}$ orbitals
[see T. Shimojima {\it et al.}, Phys. Rev. Lett. {\bf 104}, 057002 (2010)],
since this Fermi-surface unbalance 
originates in the rotational symmetry breaking property of the $(\pi,0)$
magnetic order as explained in M. Daghofer {\it et al.}, 
Phys. Rev. B {\bf 81}, 180514(R) (2010).


\bibitem{note}
Note that in the presence of external strain to detwin crystals, 
some remaining artificial anisotropy 
may incorrectly suggest that $(\pi,0)-(0,\pi)$ are not degenerate above $T_S$ in
neutron scattering, 
leading to the incorrect conclusion that $T_{PG}$ is $T_S$
(for related observations see C. Dhital {\it et al.}, 
Phys. Rev. Lett. {\bf 108}, 087001 (2012);
E. C. Blomberg {\it et al.}, 
Phys. Rev. B {\bf 85}, 144509 (2012)).



\bibitem{PG-exper} Our results should be compared against the photoemission
experiments reported by T. Shimojima {\it et al.}, Phys. Rev. B {\bf 89}, 045101 (2014)
(see for instance their Figure 6).
Infrared studies correlating the presence of a pseudogap
with antiferromagnetic fluctuations can also be found in S. J. Moon {\it et al.}, 
Phys. Rev. Lett. {\bf 109}, 027006 (2012).

\bibitem{rong} Rong Yu {\it et al.}, 
Phys. Rev. B {\bf 79}, 104510 (2009). 

\bibitem{egami} J. L. Niedziela, M. A. McGuire, and T. Egami, Phys. Rev. B {\bf 86},
174113 (2012), and references therein.

\bibitem{rosenthal} E. P. Rosenthal {\it et al.}, Nature Phys. {\bf 10}, 225 (2014).




\end{thebibliography}

\begin{thebibliography}{10}

\bibitem{PRL-2013-SM} S. Liang, 
A. Moreo, and E. Dagotto, Phys. Rev. Lett. {\bf 111}, 047004 (2013).

\bibitem{three-SM} M. Daghofer {\it et al.}, 
Phys. Rev. B {\bf 81}, 014511 (2010).


\bibitem{fisher-science-SM} J-H. Chu, H-H. Kuo, J. G. Analytis, and I. R. Fisher, Science {\bf 337}, 710 (20121); H.~H. Kuo {\it et al.}, Phys. Rev. B {\bf 88}, 085113 (2013);
and references therein.

\bibitem{kasahara-SM} S. Kasahara {\it et al.}, 
Nature {\bf 486}, 382 (2012).

\bibitem{fernandes2-SM} R. M. Fernandes and J. Schmalian,
Supercond. Sci. Technol. {\bf 25}, 084005 (2012).

\bibitem{PRL-2012-SM} 
S. Liang, G. Alvarez, C. Sen, A. Moreo, and E. Dagotto, 
Phys. Rev. Lett. {\bf 109}, 047001 (2012). 




\end{thebibliography}
\end{document}